\documentclass{aa}
\usepackage{graphicx}

\newcommand\simgt{\lower.5ex\hbox{$\;\buildrel>\over\sim\;$}}
\newcommand\simlt{\lower.5ex\hbox{$\;\buildrel<\over\sim\;$}}

\begin{document}

\title{Molecular Gas in NUclei of GAlaxies (NUGA)\thanks{Based on 
observations conducted at the IRAM Plateau de Bure Interferometer.  IRAM 
is supported by the INSU/CNRS (France), the MPG (Germany), and the IGN 
(Spain).}}

\subtitle{XII. The head-on collision in NGC\,1961}

\author{F. Combes        \inst{1}
        \and
       A.~J. Baker        \inst{2}
        \and
        E. Schinnerer        \inst{3}
        \and
        S. Garc\'{\i}a-Burillo        \inst{4}
        \and
        L.~K. Hunt        \inst{5}
        \and
        F. Boone        \inst{1}
        \and
        A. Eckart        \inst{6}
        \and
        R. Neri        \inst{7}
        \and
        L.~J. Tacconi        \inst{8}
}

\offprints{F. Combes}

\institute{Observatoire de Paris, LERMA, 61 Av. de
                l'Observatoire, 75014 Paris, France\\
                \email{francoise.combes@obspm.fr} and \email{frederic.boone@obspm.fr}
        \and
                Rutgers, The State University of NJ, 136 Frelinghuysen Road, 
                Piscataway, NJ 08854-8019, United States \\
                \email{ajbaker@physics.rutgers.edu}
        \and
                Max-Planck-Institut f\"ur Astronomie, 
                K\"onigstuhl 17, 69917 Heidelberg, Germany\\
                \email{schinner@mpia.de}
        \and
                Observatorio Astron{\' o}mico Nacional (OAN), Observatorio 
                de Madrid, Alfonso XII, 3, 28014 Madrid, Spain\\
                \email{burillo@oan.es}
        \and
                INAF-Osservatorio Astrofisico di Arcetri, Largo E. 
                Fermi 5, 50125 Firenze, Italy\\
                \email{hunt@arcetri.astro.it}
        \and
                Universit{\" a}t zu K{\" o}ln, I. Physikalisches Institut, 
                Z{\" u}lpicher Strasse 77, D-50937 K{\" o}ln, Germany\\
                \email{eckart@ph1.uni-koeln.de}
        \and
                Institut de Radio-Astronomie Millim{\' e}trique, 
                300 rue de la Piscine, 38406 St. Martin-d'H{\` e}res, France\\
                \email{neri@iram.fr}
        \and
               Max-Planck-Institut f{\" u}r extraterrestrische Physik, 
                Postfach 1312, 85741 Garching, Germany\\
                \email{linda@mpe.mpg.de}
             }

\date{Received ? ? 2009 / Accepted ? ? 2009}

\abstract{We present high--resolution maps of the CO(1--0) and CO(2--1) 
emission from the LINER 2 galaxy NGC\,1961.  This galaxy is unusual among 
late--type (Sc) disk galaxies in having a very large radial extent and
inferred dynamical mass. We propose a head-on collision scenario to
explain the perturbed morphology of this galaxy-- both the off-centered
rings and the inflated radius. This scenario is supported by the detection 
of a steep velocity gradient in the CO(1-0) map at the position of a southwest
peak in radio continuum and near-infrared emission. This peak would represent the remnant
of the disrupting companion. We use 
numerical models to demonstrate the plausibility of the scenario.
While ram pressure stripping could in principle be important for shocking
the atomic gas and produce the striking head-tail morphology,
the non detection of this small galaxy group in X-ray emission
suggests that any hot intragroup medium has too low a density.
A prediction of the collision model is the propagation of 
ring waves from the center to the outer parts, superposed on
a probable pre-existing $m=2$ barred spiral feature, accounting for
the observed complex structure of rings and spokes.  This lopsided wave accounts
for the sharp boundary observed in the atomic gas on the southern side. Through
dynamical friction, the collision finishes quickly in a minor merger, the 
best fit being for a companion with a mass ratio 1:4.
 We argue that NGC\,1961 has a strongly warped disk, which
gives the false impression of a nearly face-on system; the main
disk is actually more edge-on, and this error in the true inclination has 
led to the surprisingly high dynamical mass for a morphologically 
late-type galaxy. In addition, the outward propagating ring artificially
enlarges the disk. The collision de-stabilizes the inner disk 
and can provide gas inflow to the active nucleus.

}

\maketitle

\section{Introduction} \label{s-intro}

The tight correlation between black hole mass and central velocity dispersion 
($M_\bullet - \sigma$ relation,
Merritt \& Ferrarese \cite{merr01}; Tremaine et al. \cite{trem02}) offers 
strong evidence that galaxies' stellar spheroids and central black holes 
are built up by mechanisms that are closely linked, if not identical.  In disk 
galaxies, black holes of increasing mass can scatter more disk stars into a 
bulge (Norman et al. \cite{norm96}) and drive secular evolution along the 
Hubble sequence.  A more generally applicable scenario (holding for both early 
and late-type galaxies), however, centers on the balance of nuclear 
fuelling and feedback.  If the relative fractions of inflowing gas consumed by nuclear star 
formation and accretion onto a black hole are roughly constant, then the 
$M_\bullet - \sigma$ relation is naturally explained.  

Since nearly half of all nearby systems host  low luminosity active galactic nuclei (AGN: Ho 
et al. \cite{ho97a}), the supply of fuel to the Schwarzschild radius (not to 
mention circumnuclear star-forming regions) is still very much an ongoing 
process.  Because black holes appear to occupy the centers of nearly all 
galaxies with massive bulges, high--resolution studies of the gas dynamics in 
virtually {\it any} system with a massive spheroid should therefore shed light 
on the role of fuelling in driving the coevolution of black holes and host 
galaxies.  This prospect is the motivation for our observations of the {\bf
Nu}clei of {\bf Ga}laxies (NUGA) sample at the IRAM Plateau de Bure 
Interferometer (PdBI: Guilloteau et al. \cite{guil92}).  The NUGA survey is 
designed to explore the molecular gas content and mechanisms for nuclear 
fuelling at high angular resolution; initial observations of a sample of 
twelve nearby AGN (Garc\'{\i}a-Burillo et al. \cite{garc03}) will ultimately be 
combined with a larger sample of both active and quiescent nuclei for the 
purpose of statistical comparisons.  In this paper, the twelfth in an initial 
series focused on individual objects, we present the PdBI data for the 
NUGA target NGC\,1961.

\begin{figure}
\centering
\includegraphics[width=8.5cm]{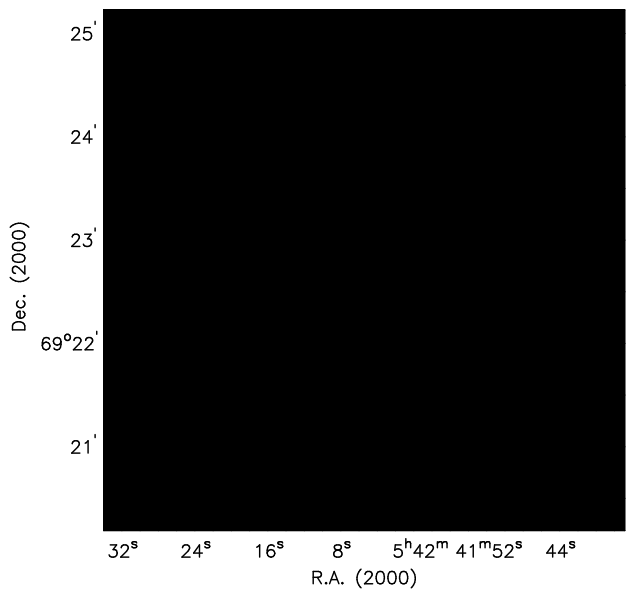}
\caption{Blue DSS image of NGC\,1961.  Note the sharp linear feature to the 
southeast (oriented at 45$^\circ$), and the anomalous arm winding from south to west.
The ``arm features'' are tightly wound, and it is not easy to identify their
winding sense. They might even wind opposite to the global sense, at some radii.
In any case, the morphology is better described by rings.}
\label{f-opt}
\end{figure}

Judged by its nuclear properties, NGC\,1961 is a relatively unremarkable 
low--luminosity AGN.  Its optical spectrum is classified as a Low-Ionization 
Nuclear Emission--line Region (LINER) by Ho et al. (\cite{ho97b}); it shows no 
broad-line emission, although its narrow-line profiles do show asymmetric blue 
wings (Ho et al. \cite{ho97c}).  NGC\,1961 has attracted significantly more 
attention for being one of the most massive known disk galaxies (Rubin et al. 
\cite{rubi79}), and moreover unusually massive for its late (Sc) Hubble type.  
This peculiarity cannot be attributed to a distance error, since NGC\,1961 is 
associated with a group of ten galaxies that all lie within $514\,{\rm 
km\,s^{-1}}$ of each other at small separations on the sky (Gottesman et al. 
\cite{gott02}).  NGC\,1961 appears to be impinging on this group from the 
northwest: its large-scale optical morphology shows a pronounced linear 
feature to the southeast which connects to an anomalous asymmetric arm winding 
from south to west (Figure \ref{f-opt}). 

CO emission has already been detected in  NGC\,1961 (Young et al.
\cite{youn95}; Komugi et al. \cite{komu08}), corresponding to a molecular
gas mass of $2.7 \times \,10^{10}\,M_\odot$.  The H\,I content of 
NGC\,1961 is substantial (nearly $5 \times 10^{10}\,M_\odot$ of atomic 
hydrogen: Haan et al. \cite{haan08}), which suggests a very late type galaxy. 
However, the galaxy has a large radial extent, both optically and in H\,I, 
apparently much larger than expected for its morphological type. The H\,I 
morphology revealed by 
recent VLA observations (Haan et al. \cite{haan08}) is striking because of its
head-tail shape, with a very sharp boundary in the south. This feature suggests
 a ram-pressure shock, although no diffuse X-ray emission has
been detected by {\it ROSAT}
 in this galaxy group (Mulchaey et al.  \cite{mulc03}). 
Shostak et al. (\cite{shos82}) first noted the sharp H\,I cut-off displaced
with respect to the stellar boundary and suggested a ram pressure shock 
 was responsible.
This linear feature of higher H\,I column density coincides with a higher 
radio continuum surface brightness at frequencies from 1.5 to 15.4\,GHz 
(Lisenfeld et al. \cite{lise98}).  

The galaxy as a whole has an unusually high radio continuum luminosity 
relative to its optical (blue) and far-infrared ($40 - 120\,\mathrm{\mu m}$) 
luminosities.  Lisenfeld et al. (\cite{lise98}) suggest 
that a second nucleus in their radio continuum map may be the remnant of an 
infalling dwarf galaxy; they attribute the high radio luminosity to
cosmic rays that have been produced in the putative minor merger. The
observed morphology can constrain its orbit type:
the lack of extended tidal tail features in the H\,I map 
suggests that any recent merger at the 
very least cannot have had a prograde encounter geometry with respect to 
NGC\,1961's sense of rotation.  We will use the molecular line observations 
and the numerical simulations reported in this paper 
to help identify which of these two mechanisms is more 
plausibly responsible for the large-scale asymmetries in NGC\,1961.

We describe our observations in \S \ref{s-obs} and their results in \S 
\ref{s-res}, focusing on the $m = 1$ asymmetry 
and off-centered ring morphology evident in the CO maps.  
A dynamical model is proposed and arguments developed in 
\S \ref{s-mod} to explain the perturbed morphology and asymmetries 
observed on large scales as the result of a recent head-on collision.  
\S \ref{simu}  describes in detail the adopted model,
the technique of the simulations, and their results. 
 Our discussion of the collision scenario and its 
implications for AGN fuelling are presented in \S \ref{discuss}, and our 
conclusions in \S \ref{s-conc}.  Throughout this paper, we will use the 
distance $D = 52.4\,\mathrm{Mpc}$ specified by Tully (\cite{tull88}), and a 
CO--to--${\rm H_2}$ conversion factor $X \equiv N_\mathrm{H_2}/
I_\mathrm{CO(1-0)} = 2.2 \times 10^{20}\,\mathrm{cm^{-2}\,(K\,km\,
s^{-1})^{-1}}$ as recommended by Solomon \& Barrett (\cite{solo91}).

\section{Observations} \label{s-obs}

\subsection{CO interferometry} \label{ss-obsco}

\begin{table}
\caption{Log of PdBI observations of NGC\,1961.  For each dataset, we list 
the configuration, the number of antennas, and the 3\,mm/1\,mm flux densities 
adopted for the gain calibrator 0716+714 (in mJy).}
\begin{tabular}{cccc}
\hline
Date        & Config & $N_{\rm ant}$ & 0716+714 \\
\hline
18 Dec 2000 & C & 4 & 1.3 / 0.7 \\
21 Dec 2000 & C & 5 & 1.0 / 0.9 \\
26 Mar 2001 & D & 5 & 0.7 / 0.6 \\
31 Mar 2001 & D & 4 & 0.5 / 0.4 \\
09 Jan 2002 & B & 6 & 1.0 / 2.1 \\
03 Mar 2003 & A & 6 & 2.0 / 1.8 \\
\hline
\end{tabular}
\label{t-obs}
\end{table}

We observed the CO(1--0) and CO(2--1) lines in NGC\,1961 with the 
IRAM PdBI between December 2000 and January 2002.  During this period, the 
array included from four to six 15\,m diameter antennas; deployment in the B, 
C, and D configurations yielded a total of 35 distinct baselines ranging from 
24\,m to 280\,m in length.  In March 2003, the A configuration
added longer baselines, up to 420\,m  for the CO(2-1) line, and provided
sub-arcsecond resolution
( see documentation on configurations at http://www.iram.fr).
Each antenna was equipped with both 
single--sideband 3\,mm and double--sideband 1\,mm SIS receivers that were used 
simultaneously; receiver temperatures were 60--80\,K, and corrected outside 
the atmosphere yielded typical system temperatures of $\sim 120\,{\rm K}$ and 
$\sim 400\,{\rm K}$ at 3\,mm and 1\,mm, respectively. We deployed four 
correlator modules-- giving a total of 580\,MHz of continuous bandwidth at 
2.5\,MHz resolution-- across each of the two lines.  The CO(1--0) and CO(2--1) 
lines were observed in the upper and lower sidebands relative to the 3\,mm 
and 1\,mm reference frequencies.  As the phase (pointing) center for the 
observations, we adopted a galaxy's optical position, slightly
different from that listed in NED-- 
$\alpha_{\rm J2000} =$ 05:42:04.8 and $\delta_{\rm J2000} =$
+69:22:43.3.\footnote{These coordinates were originally derived relative to 
the position of SN 1998eb.}

We calibrated our data using the CLIC routines in the IRAM GILDAS software 
package (Guilloteau \& Lucas \cite{guil00}).  For passband calibration, we 
used the bright quasars 3C111, 3C279, and 3C454.3.  Observations of the blazar 
0716+714 interleaved with our source data every $\sim 30$ minutes were used to 
remove phase drifts and amplitude gain variations.  The flux density of 
0716+714 was in turn determined at each epoch by comparisons with the 
 calibration
sources CRL618 and MWC349; we estimate the uncertainties in the flux scales to 
be $\sim 20\%$.  Table \ref{t-obs} lists the details of our observations and 
calibrations, including the flux densities adopted for 0716+714 at both 3\,mm 
and 1\,mm. The fact that 0716+714 is a blazar explains its strong variability.

Before constructing $uv$ tables from our calibrated data, we smoothed them to
a frequency resolution of 5\,MHz (velocity resolutions $13.2\,{\rm 
km\,s^{-1}}$ for the CO(1--0) line and $6.6\,{\rm km\,s^{-1}}$ for the 
CO(2--1) line).  
 The primary beam sizes at the two frequencies are 43 and 21.5\arcsec
respectively.
The full datasets include the (on--source, six--telescope) 
equivalents of 16.8 hours of 3\,mm data and 13.9 hours of 1\,mm data.  In 
the 3\,mm data cube, CO(1--0) line emission is detected out to $330\,{\rm 
km\,s^{-1}}$ blueward and $369\,{\rm km\,s^{-1}}$ redward of the systemic 
$z_{\rm CO} = 0.01331 \sim v_{\rm sys} = 3937\,{\rm km\,s^{-1}}$ LSR (see \S 
\ref{ss-cokin}).  We constructed a 3\,mm continuum channel from the 
upper--sideband channels outside this velocity range, and upon mapping (with 
natural weighting) discovered the presence of a $1.3 \pm 0.2\,{\rm mJy}$ 
continuum source offset $\sim 3\farcs4$ to the northwest of the nucleus (see 
\S \ref{ss-ctmres}).  In order to avoid contamination of the CO(1--0) line 
maps, we used the GILDAS task UV\_SUBTRACT to remove a $uv$ model for this 
continuum source from all of the 3\,mm $uv$ data.  Our maps of the 
continuum--subtracted line cube used natural weighting, which (as for the 3\,mm
continuum data) yielded a synthesized beam of $2\farcs73 \times 2\farcs41$ 
at position angle $36^\circ$.  The rms per $13.2\,{\rm km\,s^{-1}}$ channel is 
$1.3\,{\rm mJy\,beam^{-1}}$.

At 1\,mm, no continuum emission was evident  (with an rms of 0.4 mJy)
in a map made from the upper 
(line--free) sideband visibility data, leading us to conclude that no 
subtraction of continuum from our 1\,mm line data would be necessary.  
CO(2--1) emission is seen over a velocity range of $\pm 257\,{\rm km\,s^{-1}}$ 
relative to $v_{\rm sys}$, smaller than the full CO(1--0) range because gas 
moving at the highest velocities in the CO(1--0) data cube lies outside the 
smaller CO(2--1) primary beam.  We used uniform weighting to reach an angular 
resolution of $0\farcs91 \times 0.\farcs84$ ($\sim 231\,{\rm pc} \times 
214\,{\rm pc}$ in projection) at position angle $24^\circ$.  The rms per 
$6.6\,{\rm km\,s^{-1}}$ channel is $4.3\,{\rm mJy\,beam^{-1}}$.

We measured the integrated intensity ratio ${\cal R}_{21/10} = I_{\rm CO(2-1)}/
I_{\rm CO(1-0)}$ as a function of position using maps made from $uv$ datasets 
truncated at an identical radius of $B/\lambda = 1.6 \times 10^4$ for 
both lines (i.e., slightly larger than the minimum $uv$ radius reached for the 
CO(2--1) data on the shortest projected baselines).  The loss of 
short--spacing information means that a map made from the truncated CO(1--0) 
data has a smaller synthesized beam 
than the original version made from the full data; the 
new CO(2--1) map, in contrast, is hardly changed at all.  We then applied 
corrections for the primary beam response to both maps, convolved the CO(2--1) 
map to the resolution of the ($uv$-truncated) CO(1--0) map, and divided them.

\begin{table}[ht]
\caption{ Parameters for NGC\,1961 \label{param} }
\begin{flushleft}
\begin{tabular}{lll}
\hline
Parameter  &   Value                              &  Reference \\
\hline
R.A. (J2000)  &   05$^{\rm h}42^{\rm m}04.6^{\rm s}$ & NED \\
Dec. (J2000) &   69$^\circ$22$\arcmin$42.4\arcsec    & NED \\
$v_{\rm hel}$ &   $3934\,{\rm km\,s^{-1}}$           & NED \\
RC3 Type   &    SAB(rs)c                             & RC3 \\
Diameters &         4.6\arcmin $\times$ 3\arcmin     & LEDA  \\
Inclination &         47$^\circ$                  & LEDA  \\
Position angle &         85$^\circ$               & LEDA  \\
Distance    &         52.4 Mpc (1\arcsec= 254\,pc)    & T88 \\
$M$(H\,I)                 &  $4.7 \times 10^{10}\,M_\odot$    & H08 \\
$L_B$       &    $8.9 \times 10^{10}\,L_{\odot}$        & LEDA \\
$L_{\rm FIR}(40-120\,{\rm \mu m})$  & $10.7 \times 10^{10}\,L_{\odot}$  
& {\it IRAS} \\
R.A. (J2000)  &   05$^{\rm h}42^{\rm m}04.7^{\rm s}$ & CO-dyn \\
Dec. (J2000) &   69$^\circ$22$\arcmin$42.3\arcsec    & CO-dyn \\
\hline
\end{tabular}
\end{flushleft}
RC3: de~Vaucouleurs et al. (\cite{deva91}) \\
T88: Tully (\cite{tull88}) \\
H08: Haan et al. (\cite{haan08}) \\
CO-dyn: Dynamical center derived from the CO(2-1) kinematics
\end{table}

\subsection{Optical and infrared imaging} \label{ss-obsopt}

We have extracted optical and near-infrared images of NGC\,1961 from the 
{\it Hubble Space Telescope} ({\it HST}) archive.  Images were acquired using 
WFPC2 through the F547M and F606W ($\sim V$) filters, and using NICMOS through 
the F160W ($\sim H$) filter (see 
Martini et al.  \cite{mart03}, Hunt \& Malkan \cite{hunt04}).  
To show the large-scale 
optical morphology of the galaxy, we have taken red and blue optical images 
from the Digitized Palomar Sky Survey (DSS); the blue image is shown in Figure 
\ref{f-opt}.

For wide-field near-IR information, we have turned to the Two Micron All Sky
Survey (2MASS: Skrutskie et al. \cite{skru97}).  Jarrett et al. 
(\cite{jarr03}) have included NGC\,1961 in their 2MASS ``Large Galaxy Atlas'';
we have downloaded their reduced images from the IPAC website.  A $K_{\rm
  s}$ image is shown in Section \ref{s-mod}.

IRAC images at 3.6\,$\mu$m,   4.5\,$\mu$m and 8.0\,$\mu$m
were retrieved from the Spitzer archive.
We started with the Basic Calibrated Data images,
and aligned and combined them with MOPEX 
(Makovoz \& Marleau \cite{mako05}) which accounts
for distortion and rotates to a fiducial coordinate system.
1\farcs20 pixels were imposed for the final images,
roughly the same as the original IRAC detector.
Significant banding was present %from a bright star in the field,
and this was corrected for by
interpolation over the affected rows before combination with MOPEX.
 The 8.0\,$\mu$m IRAC filter tends to be dominated by aromatic features of 
polycyclic aromatic hydrocarbons (PAHs), but also contains a contribution 
from stellar photospheres. The IRAC images at shorter wavelengths (3.6, 
4.5\,$\mu$m) can be suitably combined and scaled to estimate the stellar 
emission. Following Helou et al. (2004), we obtained a non-stellar (dust) 
image by subtracting
from the 8.0\,$\mu$m image a scaled linear combination of the 3.6
and 4.5\,$\mu$m images. The IRAC images were originally
published by Pahre et al. (\cite{pahr04}).

\section{Results} \label{s-res}

\subsection{CO line morphologies} \label{ss-comorph}

\begin{figure}
\centering
\includegraphics[width=8.5cm]{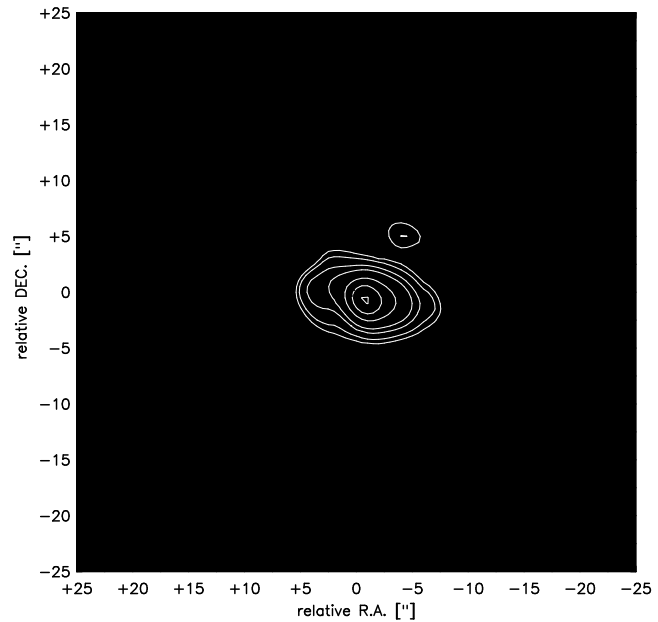}
\caption{CO(1--0) zeroth moment map, with coordinates defined relative to the 
PdBI phase center.  The $2.73\arcsec \times 2.41\arcsec$ synthesized beam is 
plotted at lower left; contours are \{0.04, 0.08, 0.12, 0.16, 0.2, 0.3, 0.4, 
0.5, 0.6, 0.8, 1.0, 1.5, 2.0, 2.5\} $\times 13.2\,{\rm Jy\,beam^{-1}\,km\,
s^{-1}}$.}
\label{f-co10m0}
\end{figure}

\begin{figure}
\centering
\includegraphics[width=8.5cm]{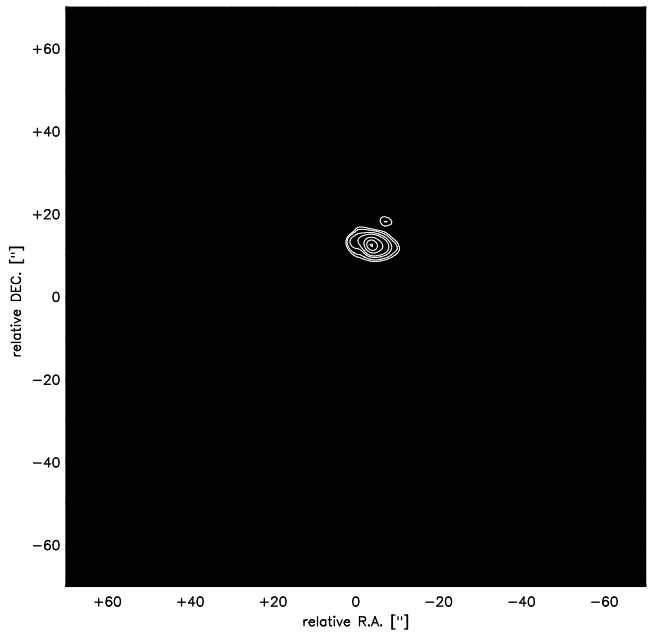}
\caption{CO(1--0) zeroth moment map contours (as in Figure \ref{f-co10m0}) 
overlaid on the 8.4\,GHz radio continuum map from Lisenfeld et al. 
(\cite{lise98}) in greyscale.  Coordinates are defined relative to the VLA 
phase center adopted for the 8.4\,GHz observations, whose $\sim 8\arcsec$ 
synthesized beam is plotted at lower left.  CO(1--0) emission coincides with 
the two strongest 8.4\,GHz peaks, one in the nucleus of NGC\,1961 and the 
other offset to the southwest.}
\label{f-co10+8.4}
\end{figure}

\begin{figure}
\centering
\includegraphics[width=8.5cm]{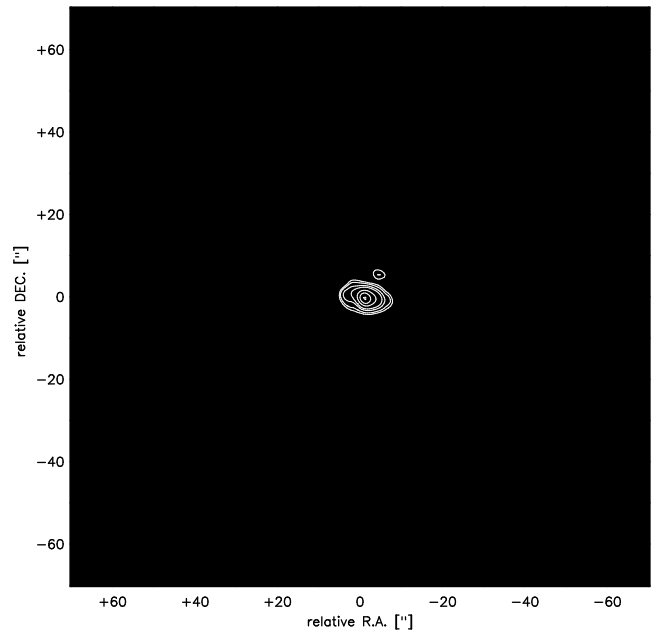}
\caption{PdBI CO(1--0) zeroth moment map contours (as in Figure 
\ref{f-co10m0}) overlaid on the blue DSS image in greyscale.  Coordinates are 
defined relative to the PdBI phase center.}
\label{f-co10+dss}
\end{figure}

We derived zeroth moment maps from our CO data cubes using the Groningen Image 
Processing SYstem (GIPSY: van~der~Hulst et al. \cite{vand92}); emission was 
integrated over the observed line widths after clipping all $\leq 3\sigma$ 
pixels in the individual channels.  Figure \ref{f-co10m0} shows the result for 
the CO(1--0) line.  The global molecular gas distribution is dominated by a 
strong central concentration of dimensions $\sim 10\arcsec \times 6\arcsec$.  
This structure is encircled to the north and east by a winding gaseous arc 
containing a number of knots, the strongest of which lie $\sim 15\arcsec$ east
and $\sim 7\arcsec$ northwest of the nucleus.  To the southwest, there is an 
isolated concentration of gas whose relationship to the rest of the disk is 
not immediately apparent.  An important clue to its nature comes from Figure 
\ref{f-co10+8.4}, which shows CO(1--0) contours overlaid on a greyscale image 
of 8.4\,GHz continuum emission from Lisenfeld et al. (\cite{lise98}).  
Although the CO(1--0) primary beam does not extend over the entire 8.4\,GHz 
map, a comparison in the inner $\sim 1\arcmin$ is already instructive.  The 
anomalous CO(1--0) emission southwest of the nucleus coincides almost exactly 
with the off--center 8.4\,GHz feature that Lisenfeld et al. (\cite{lise98}) 
identify as a ``second nucleus.''  This agreement adds some weight to the 
authors' suggestion that the off--nuclear peak represents the remnant of a 
smaller galaxy that has merged with NGC\,1961 (see also \S \ref{s-mod} below).
 There is a slight offset between the two positions, which is certainly 
expected, since the small companion is likely to be strongly perturbed and the 
CO emission itself is lopsided.

In Figure \ref{f-co10+dss}, we show an overlay of the CO(1--0) contours of 
Figure \ref{f-co10m0} on the (inner region of the) blue DSS image of Figure 
\ref{f-opt} in greyscale.  The anomalous CO(1--0) feature to the southwest of 
the nucleus coincides with a region of high extinction in the optical image, 
suggesting that the molecular gas structure may lie on the near side of the 
galactic disk.  This region also corresponds to the possible second nucleus.
The eastern section of the winding arc also aligns very nicely 
with a region of high extinction, very likely a dust lane, although the 
northern section does not.  This contrast offers the first of several hints 
that the CO(1--0) arc does not trace a single spiral arm.
 Instead, as will be clear in the CO(2--1) map, the CO emission suggests
 a distorted ring-like structure. 

The reliability of the CO(1--0) line luminosity as an extragalactic mass 
tracer (e.g., Dickman et al. \cite{dick86}; Solomon \& Barrett \cite{solo91}) 
allows us to convert the line fluxes of various structures in the CO(1--0) map 
 into molecular gas masses.  Including a factor of 1.36 to account for helium, 
the gas mass bound into molecular clouds can be calculated as
\begin{eqnarray}
{\frac {M_{\rm gas}}{M_\odot}} & = & 3.23 \times 10^7\,\Big({\frac {F_{\rm 
CO(1-0)}}{\rm Jy\,km\,s^{-1}}}\Big) \nonumber \\
 & & \times \Big({\frac X {2.2 \times 10^{20}}}\Big)\, \times \Big({\frac D {\rm 
52.4\,Mpc}}\Big)^2
\end{eqnarray}
After correcting Figure \ref{f-co10m0} for primary beam response, we estimate 
the masses of the central disk, and of the CO ring including the anomalous 
southwestern emission, to be $1.4 \times 10^{10}\,M_\odot$
 and $1.1 \times 10^{10}\,M_\odot$, respectively.  The molecular mass corresponding
to the southwestern emission alone is $1.5 \times 10^{9}\,M_\odot$. Summing the 
CO(1--0) line flux over the galaxy's observed area gives a total molecular mass 
of $M_{\rm gas} = 2.5 \times 10^{10}\,M_\odot$.
This molecular component is more than an order of magnitude
 more massive than in any NUGA galaxy studied so far.

\begin{figure}
\centering
\includegraphics[width=8.5cm]{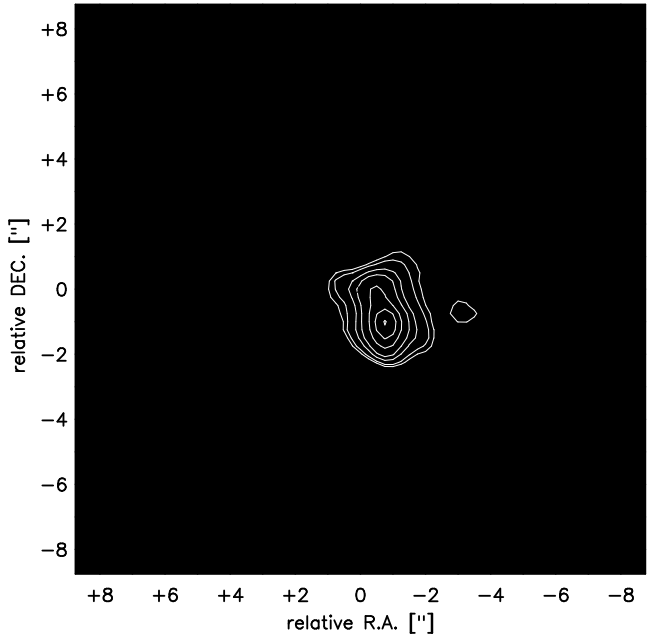}
\caption{CO(2--1) zeroth moment map, with coordinates defined relative to the 
PdBI phase center.  The $0.91\arcsec \times 0.84\arcsec$ synthesized beam is 
plotted at lower left; contours are \{0.12, 0.2, 0.3, 0.4, 0.5, 0.6, 0.8, 1.0, 
1.5, 2.0, 3.0, 4.0, 5.0\} $\times 6.6\,{\rm Jy\,beam^{-1}\,km\,s^{-1}}$.}
\label{f-co21m0}
\end{figure}

\begin{figure}
\centering
\includegraphics[width=8.5cm]{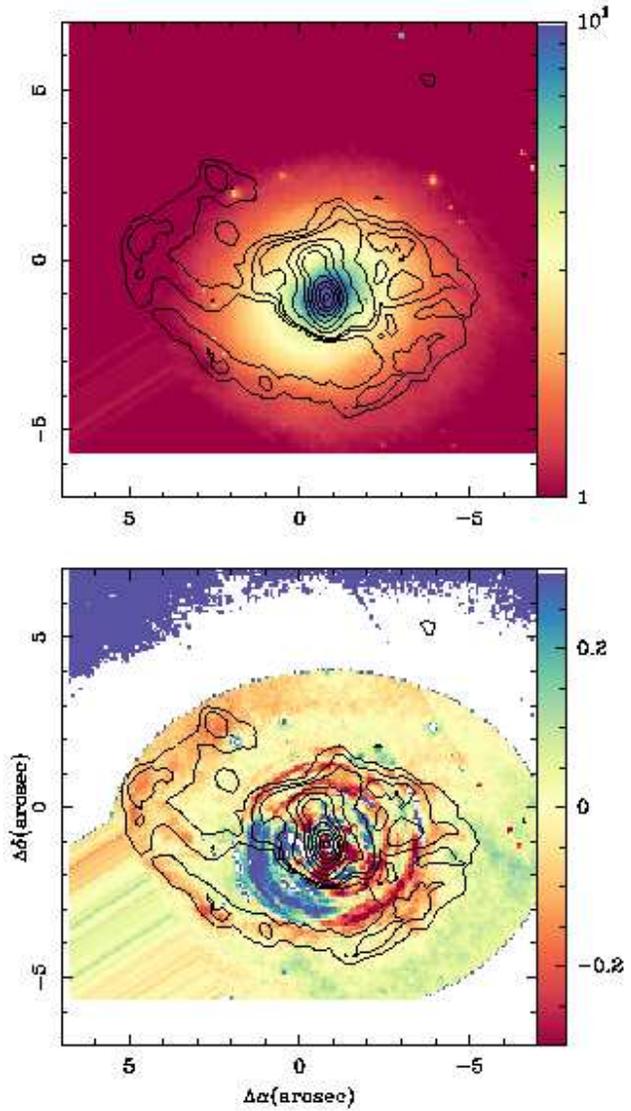}
\caption{{\bf Top} CO(2--1) zeroth moment contours (as in Figure \ref{f-co21m0}) 
overlaid on {\it HST} F160W colourscale.  {\bf Bottom} Same contours superposed
on the Unsharp Mask version of the NICMOS image.}
\label{f-co21+h}
\end{figure}

\begin{figure}
\centering
\includegraphics[width=8.5cm]{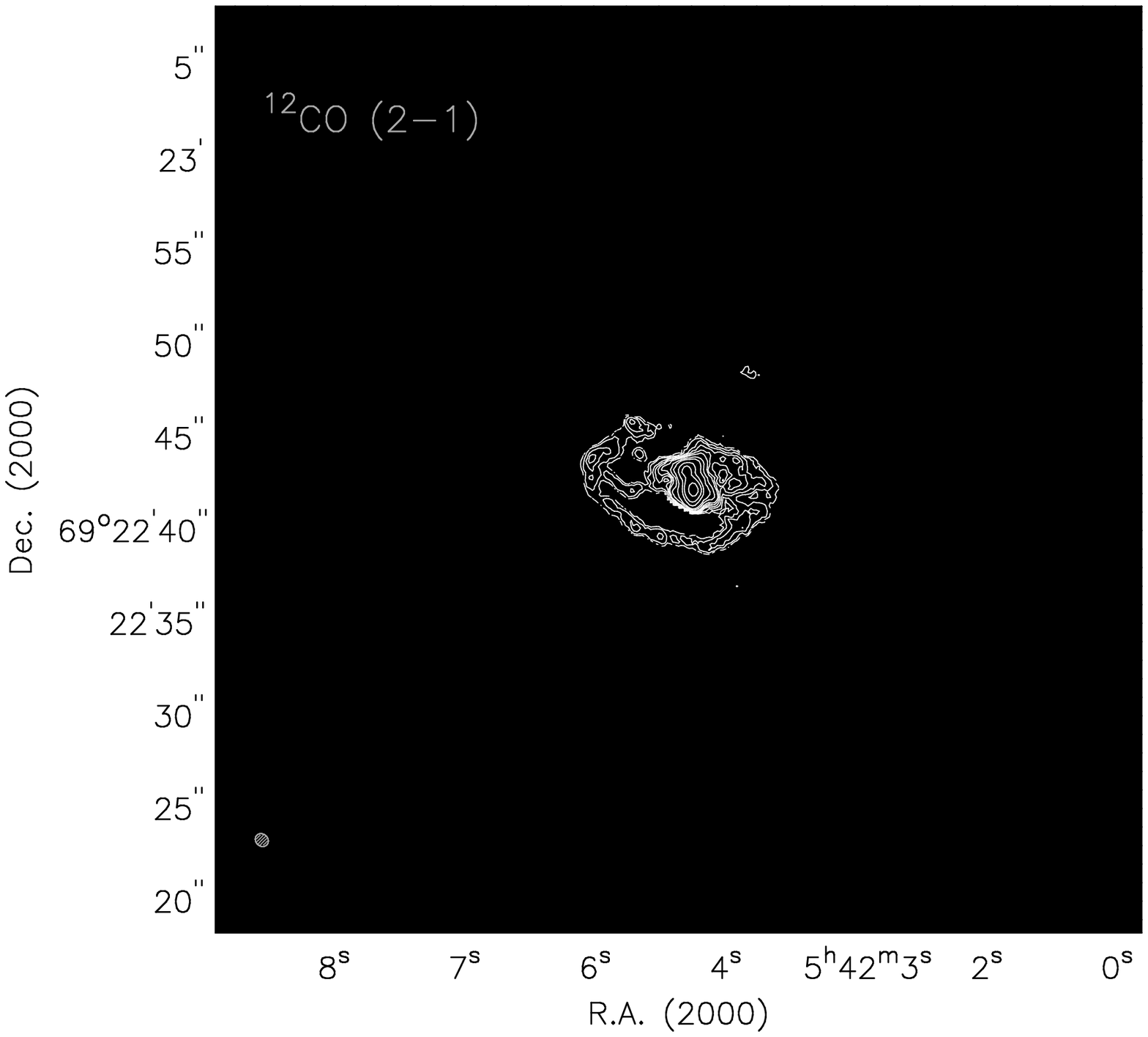}
\caption{CO(2--1) zeroth moment contours (as in Figure \ref{f-co10m0}) 
overlaid on CO(1--0) zeroth moment greyscale.}
\label{f-co21+co10}
\end{figure}

Figure \ref{f-co21m0} shows the zeroth moment of the CO(2--1) line.  Due to 
primary beam attenuation, only the strongest ($\sim 7\arcsec$ northwest) knot 
in the northern portion of the CO(1--0) arc is detected in CO(2--1).  However, 
the central concentration in the CO(1--0) map is clearly resolved into a more 
compact structure encircled to the {\it south} by a clearly continuous but 
strongly asymmetric spiral arm or arc.  This southern arm coincides with an optical 
dust lane, which is better seen on the CO(2--1)/unsharp mask NICMOS image overlay shown in Figure 
\ref{f-co21+h}. The southern arc winds up toward the north in a ring,
and lacks the constant-sign pitch angle that an arm would have. 
This overlay has been done by making the dynamical
CO(2--1) map coincide with the peak in the NICMOS image
(corresponding also to the peak in the CO(2--1) image).
The NICMOS image in Fig \ref{f-co21+h} (top) shows  some evidence of a nuclear
stellar bar of diameter roughly 1 kpc ($\sim$4\,arcsec), oriented along PA = - 50$^\circ$.
The signature of the nuclear bar is clearly seen in the F160W radial
brightness profile (not shown) as an ellipticy peak over a constant
PA $\simeq -50^\circ$.
This is clearly a nuclear bar, since its elongation and position angle is 
different from the rest of the disk. 

 In Figure \ref{f-co21+co10}, which shows CO(2--1) contours 
overlaid on CO(1--0) greyscale, it appears that the tip of the southern arm 
in the CO(2--1) map is more or less continuous with the northern section of 
the CO(1--0) arc.  It is obvious, however, that the complex structure of arcs 
does not correspond to a simple and classical $m=2$ spiral structure.
Instead, the pitch angle of the arcs is very low, and they wind up over
almost a turn, more closely resembling a ring morphology. 
 The CO is distributed in 2 or 3 embedded, offset rings, 
 suggesting an $m=1$ asymmetry.
 This ringed CO(2--1) + CO(1--0)  structure would then continue
to be embedded in another ring structure seen on large scales in the optical.
The external ring is much more asymmetrical and off-centered. These features 
provide 
strong clues as to the nature of the event that produced the large--scale asymmetries 
summarized in \S \ref{s-intro}. Although it is possible to explain the small-scale lopsideness
by internal mechanisms only (e.g., Jog \& Combes \cite{jog09}), the external 
ring's off-centering suggests a galaxy collision.

The innermost CO(2--1) contours reveal a double--lobed structure of dimensions 
$\sim 3\arcsec \times 2\arcsec$ that is not aligned with the near-infrared nuclear bar 
 in the {\it HST}-NICMOS image of Figure \ref{f-co21+h}.  
This structure could  correspond to the gas response to the  nuclear bar,  
although this is not the only possible explanation in all 
cases (Schinnerer et al. \cite{schi00}; Baker \& Scoville \cite{bake98}).  
The peak in the CO(2--1)
emission occurs in the southern lobe at an offset of ($-$1\arcsec, $-$1\arcsec) 
with respect to the PdBI pointing position
(Figure \ref{f-co21m0}),  and coincides with the NED optical center
(see Table \ref{param}).

\subsection{CO line kinematics} \label{ss-cokin}

\begin{figure}
\centering
\includegraphics[width=8.5cm]{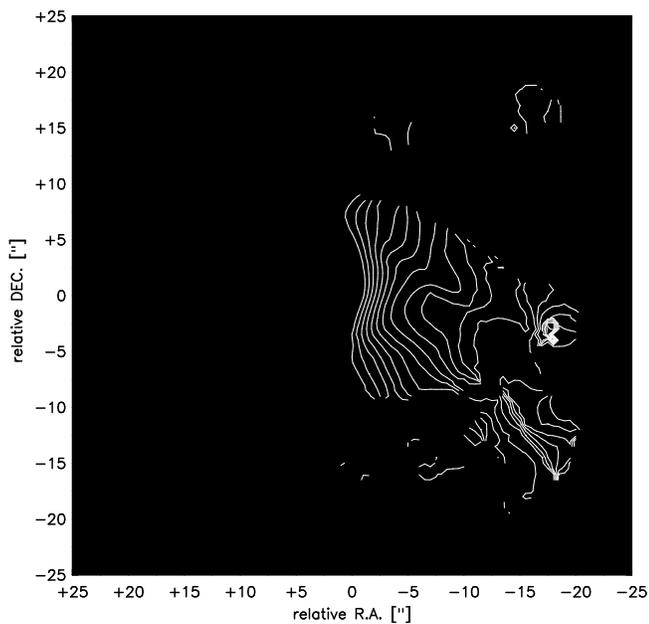}
\caption{CO(1--0) first moment map, with coordinates defined relative to the
PdBI phase center.  The synthesized beam at lower left is as in Figure 
\ref{f-co10m0}; contours are in steps of $20\,{\rm km\,s^{-1}}$ relative to 
$v_{\rm sys}$, with negative (positive) velocities indicated by black (white) 
curves.}
\label{f-co10m1}
\end{figure}

\begin{figure}
\centering
\includegraphics[width=8.5cm]{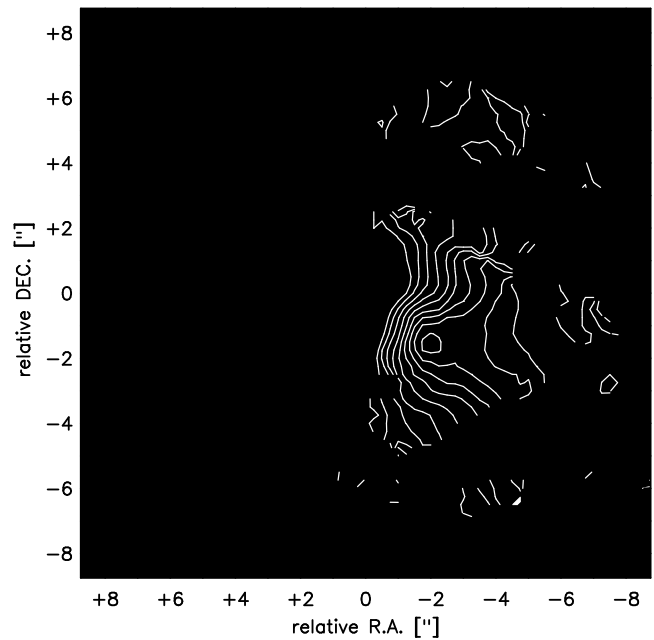}
\caption{CO(2--1) first moment map, with coordinates defined relative to the 
PdBI phase center.  The synthesized beam at lower left is as in Figure 
\ref{f-co21m0}; contours are as in Figure \ref{f-co10m1}.}
\label{f-co21m1}
\end{figure}

We have also used GIPSY to analyze the CO velocity fields in NGC\,1961.  
Figures \ref{f-co10m1} and \ref{f-co21m1} show the first moments of the 
CO(1--0) and CO(2--1) lines, constructed with clipping of (respectively) $\leq 
5\sigma$ and $\leq 3\sigma$ pixels in the original velocity channels.  As 
expected given the orientation of the galaxy's isophotal major axis on the 
sky, the predominant velocity gradient in both first moment maps runs in the 
same east--west direction; the ascending node has a position angle $\sim 
263^\circ$.  The larger amount of dust seen towards  the southern part of the disk, 
both in the DSS image on large scales and in the {\it HST} image on small scales, 
establishes that the near side is south and the far side is north.
The receding velocities in the western half of NGC\,1961 then imply  
that most arcs are trailing, but some are also leading.
As noticed in the previous section, many arcs are part of large-scale rings, with almost
zero pitch angle. When spokes are seen between two rings,
these are unambiguously trailing.

 In contrast to the fairly regular velocity field in the main 
disk, however, Figure \ref{f-co10m1} shows 
an obvious discontinuity to the southwest of the 
nucleus  at the location of the putative second 
nucleus.  If considered as a distinct structure, the velocity gradient of 
$\sim 200\,{\rm km\,s^{-1}}$ along a kinematic major axis of $\sim 135^\circ$ 
would imply an enclosed dynamical mass of $\sim 3.0 \times 10^9\,M_\odot$ 
within a radius of $5\arcsec \sim 1.3\,{\rm kpc}$,
assuming an edge-on system.  This is 3 times smaller
than the central molecular gas mass estimated in \S \ref{ss-comorph} above. 
However,  the dynamical mass could be higher,  if a more
face-on inclination is assumed.
Because the range of velocities seen in the southwestern peak is reasonably 
continuous with those seen in the adjoining region of the disk, we cannot 
completely rule out the possibility that it is in the disk plane, but 
experiencing rather strong and irregular streaming motions.
 In Figure \ref{f-co10m1},  two other kinematically perturbed regions can be seen at
$\Delta \alpha, \Delta \delta)\,=\,(-17\arcsec, -2\farcs5)$
and 
$\Delta \alpha, \Delta \delta)\,=\,(12\farcs5, 0\arcsec)$,
with a 100 km\,s$^{-1}$ gradient over 3\arcsec.
Both correspond to large molecular condensations,
that provide their own self-gravity,
in addition to the large-scale velocity gradient. These do not
require extra mass.

\begin{figure}
\centering
\includegraphics[width=8.5cm, clip=true]{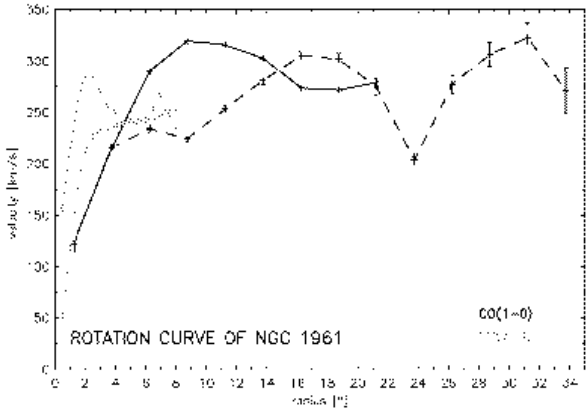}
\caption{CO--derived rotation curves for NGC\,1961.  CO(1--0) and CO(2--1) 
curves are plotted in black and gray, respectively; fits to the 
two--dimensional velocity fields with ROTCUR and to the three--dimensional 
data cubes with INSPECTOR are plotted as dashed and solid lines, 
respectively.  None of the curves are corrected for inclination.}
\label{f-corot}
\end{figure}

At the higher resolution afforded by Figure \ref{f-co21m1}, it is clear that 
the true dynamical center of NGC\,1961 lies between the two lobes seen in 
Figure \ref{f-co21m0}, rather than coincident with either of them, with the 
optical peak, or with the coordinates we adopted as the phase center for the 
PdBI observations.  A fit to the CO(2--1) velocity field locates the dynamical 
center $\sim 1\arcsec$ south and $\sim 0.5\arcsec$ west of the phase center, 
i.e., at a position $\alpha_{\rm J2000} =$ 05:42:04.7 and $\delta_{\rm J2000} 
=$ +69:22:42.3.  With respect to this center, we have derived circular 
velocity curves for NGC\,1961 in both CO lines.  Using the GIPSY task ROTCUR 
in automated mode, we have derived fits to the CO(1--0) and CO(2--1) velocity 
fields depicted in Figures \ref{f-co10m1} and \ref{f-co21m1}; the results are 
presented as dashed lines in Figure \ref{f-corot}.  The effects of beam 
smearing in reducing the $dv/dr$ slope of the CO(1--0) rotation curve relative 
to the higher--resolution CO(2--1) data are immediately apparent.  As an 
alternative, we have also used the interactive GIPSY task INSPECTOR to derive 
rotation curves from the full data cubes.  These results are shown as solid 
lines in Figure \ref{f-corot}.  Because the fits to the full data cubes are 
less strongly affected by beam smearing, they yield higher peak velocities at 
small radii, and thus convey a more accurate impression of the high mass 
densities in the galaxy's inner arcseconds.

\begin{figure}
\centering
\includegraphics[width=8.5cm]{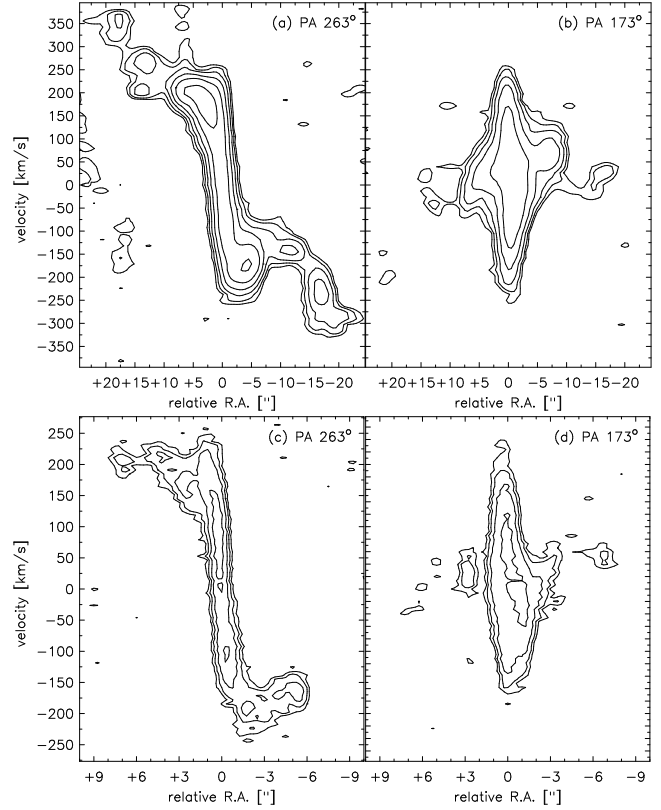}
\caption{Position--velocity cuts along the kinematic major (P.~A. $\sim 
263^\circ$, West is left) and minor (P.~A. $\sim 173^\circ$, South is left) axes of the CO(1--0) (a \& b) 
and CO(2--1) (c \& d) data cubes.  Contours are in steps of \{3, 6, 12, 
24,...\} $\times 1\sigma$.}
\label{f-copv}
\end{figure}

The full complexity of the velocity field in NGC\,1961 can perhaps best be 
appreciated from inspection of the four panels of Figure \ref{f-copv}.  Here 
we have extracted position--velocity cuts along the kinematic major (position 
angle $263^\circ$) and minor (position angle $173^\circ$) axes.  The spatial 
widths of the cuts are $50\arcsec$ for the CO(1--0) cube and $20\arcsec$ for the 
CO(2--1) cube.  Along the minor axis, we see substantial central velocity 
dispersions and indications of streaming motions within the inner $3\arcsec 
\sim 763\,{\rm pc}$.  Along the major axis, the CO(1--0) cut reveals that 
both the western section of the arc (at the most positive X-offsets) 
and-- especially-- the eastern concentration (at the most negative offsets) 
 are moving at velocities faster than would be expected if motions 
at smaller radii were straightforwardly extrapolated.  The CO(2--1) major--axis
cut is markedly asymmetric and turns over towards the east.  This pattern 
can be explained either by strong streaming motions in the central $\sim 
6\arcsec$, or by a scenario in which the eastern CO(1--0) concentration 
has its own sense of motion and is not following circular rotation in the 
disk plane.

\begin{figure}
\centering
\includegraphics[angle=-90,width=8.5cm]{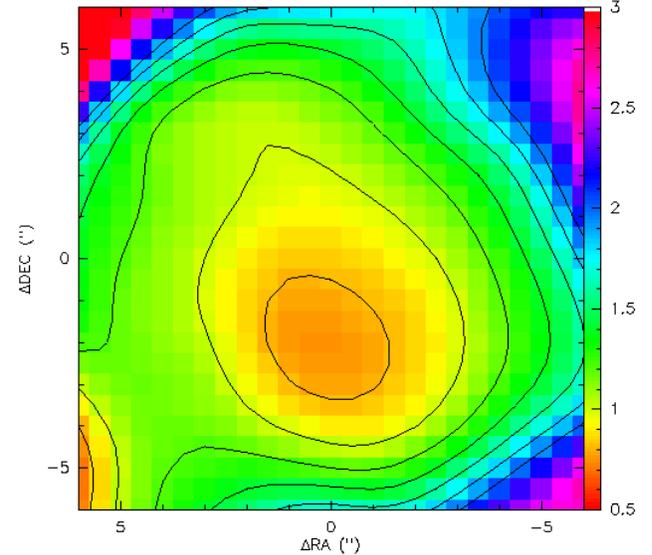}
\caption{Map of the CO(2--1)/CO(1--0) ratio obtained after smoothing
the CO(2--1) map to the resolution of the CO(1--0) map. The contours are from
0.8 to 2.0 in steps of 0.2. The ratio increases from 0.7 in the center to 
1.5 in the ring.}
\label{ratio}
\end{figure}

\subsection{CO(2--1)/CO(1--0) intensity ratio} \label{ss-corat}

Figure \ref{ratio} shows our measurement of the intensity ratio 
${\cal R}_{21/10}$ for 
regions where both CO(1--0) and CO(2--1) maps have surface brightnesses 
$\geq 2\sigma$.  Over most of the disk of NGC\,1961, we measure ${\cal 
R}_{21/10}$ to be $0.7 - 1.0$, close to the mean ratio $0.89 \pm 0.06$ measured for 
local galaxies (Braine \& Combes \cite{brai92}) and consistent with optically 
thick emission from thermally excited lines.  However, contrary to most galaxies,
where the ratio decreases from the center outwards, here the ratio increases
from 0.7 in the very center to 1.5 in the CO(2--1) ring of radius 5 \arcsec = 1.3\,kpc.
The high excitation in the ring could be due to more intense star formation.

\subsection{Millimeter continuum emission} \label{ss-ctmres}

\begin{figure}
\centering
\includegraphics[width=8.5cm]{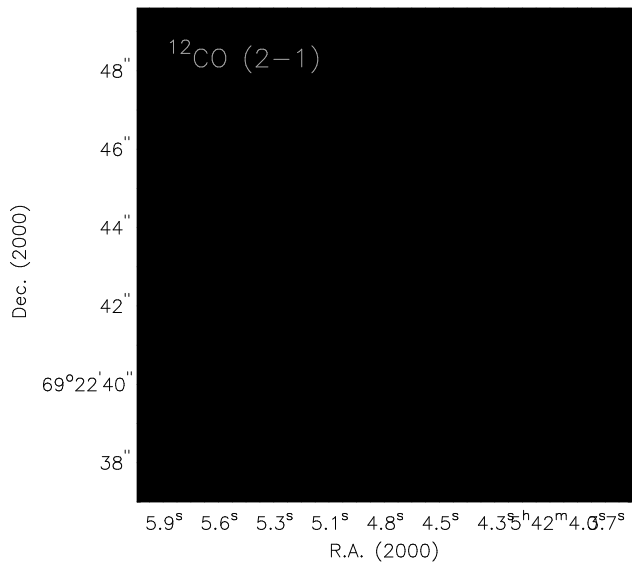}
\caption{3\,mm continumm contours overlaid on the CO(2--1) zeroth moment map 
in greyscale. The resolution of the 
3\,mm continuum map is $2.73\arcsec \times 2.41\arcsec$ at position angle 
$36^\circ$.  Contours are at 0.6 and $0.7\,{\rm mJy\,beam^{-1}}$, starting at 
3$\sigma$ (rms is $0.2\,{\rm mJy\,beam^{-1}}$).  The  dynamical center derived 
from the CO(2--1) velocity field in \S \ref{ss-cokin} is at RA=
  05$^{\rm h}42^{\rm m}04.7^{\rm s}$ and
Dec=  69$^\circ$22$\arcmin$42.3\arcsec .}
\label{f-co21on3mm}
\end{figure}

NGC\,1961 is one of a growing number of nearby AGN which reveal continuum 
emission at millimeter wavelengths (other examples include NGC\,1068, 
NGC\,3031, NGC\,3147, NGC\,3718 and NGC\,4579: Helfer et al. \cite{helf03}, Krips et al. 
\cite{krip05}, \cite{krip06}).  Our non--detection of 1\,mm continuum emission 
 (with rms of 0.4 mJy) excludes dust as a possible explanation, except in 
the unlikely event that the $0.91\arcsec \times 0.84\arcsec$ beam at 1\,mm has 
resolved out a structure that was unresolved at the $2.73\arcsec \times 
2.41\arcsec$ resolution of our 3\,mm map.  
 If the 1.3 mJy source detected at 3mm was coming from dust emission,
given typical dust temperatures,
a 1mm intensity of at least 5 mJy would be expected, 
which would have been detected at 12$\sigma$.
A flat spectrum can also be ruled out at 3$\sigma$.
We conclude that the continuum 
emission must be nonthermal in origin, implying that the source of ionizing 
photons responsible for the galaxy's optical LINER spectrum is more likely to 
be accretion than star formation.  Figure \ref{f-co21on3mm} illustrates 
the $\sim 3.4\arcsec$ offset of the continuum peak to the northeast of the 
dynamical center estimated from our fit to the CO(2--1) velocity field in \S 
\ref{ss-cokin}.  Although the $\sim 8\arcsec$ resolution of the Lisenfeld et 
al. (\cite{lise98}) radio continuum data are not sufficient to permit a 
detailed comparison, we believe this discrepancy is real; however, we do not 
view it as evidence for the migration of a supermassive black hole away from 
the center of the galaxy's potential.  Instead, as is also the case for much 
of the 3\,mm continuum emission in NGC\,1068 (Schinnerer et al. 
\cite{schi00}), we might be seeing offset jet emission.  
The implied emergence of the jet from the 
nucleus at an angle $\theta < 90^\circ$ with respect to the disk plane would 
not be exceptional, given the skewed configurations inferred for many other 
nearby low--luminosity AGN (Kinney et al. \cite{kinn00}).  
At very small scales, 
centimeter VLBI observations provide evidence  of extended 
emission, which could be the base of the jet (Krips et al. \cite{krip07}). 

\begin{figure}
\centering
\includegraphics[width=8.5cm]{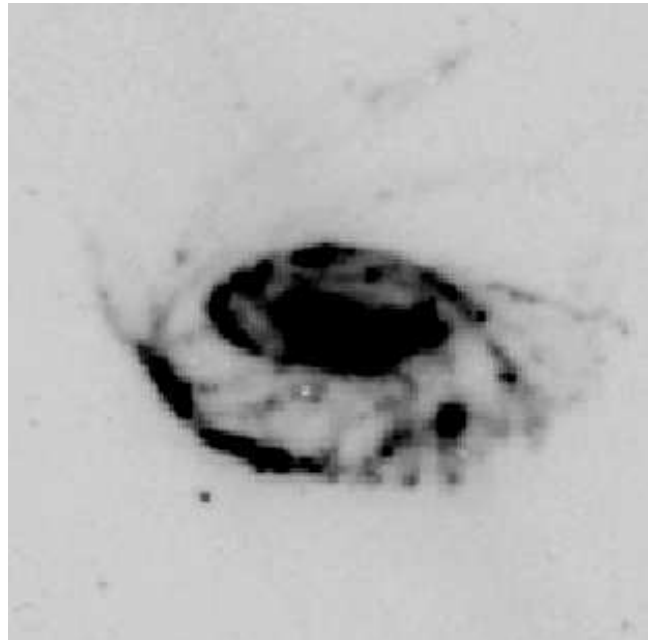}
\caption{Short-wavelength dust (non-stellar) emission from the Spitzer IRAC
images (see Sect. 2.2). North is up, East to the left; the field-of-view
is 4.2 arcmin on a side.
}
\label{nonstar}
\end{figure}

\section{A possible minor merger}
\label{s-mod}

NGC\,1961 is the largest galaxy in a small group that contains only spirals, 
with a velocity dispersion of $182\,{\rm km\,s^{-1}}$ too low for the group 
to contain hot and diffuse X-ray emitting gas
(Pence \& Rots \cite{penc97}).  The absence of an intragroup medium (IGM) 
does not favor a scenario in which ram pressure has a large effect on the 
perturbed morphology. On the other hand, 
there are no close companions presently near NGC\,1961, although
there are several group galaxies at distance larger than 100 kpc, mainly
to the south-east (Shostak et al.  \cite{shos82}).
The only plausible scenario involving a galaxy interaction 
is thus a minor merger, and there are indeed signs of a possible satellite
remnant within the confines of NGC\,1961.

The morphological and kinematic information we have gleaned from our CO(1--0) 
and CO(2--1) observations of NGC\,1961 has provided a set of important clues 
which we can use, in conjunction with data at other wavelengths,
 to try to understand the galaxy's recent dynamical history:
\begin{enumerate}

\item Most of the molecular gas within a radius of $20\arcsec \sim 5.1\,{\rm 
kpc}$ is distributed in a disk whose velocity field is reasonably regular.
At a radius of $21.5\arcsec \sim 5.5\,{\rm kpc}$, however, we see an isolated 
structure with $M_{\rm gas} \sim 1.5 \times 10^9 \,M_\odot$ and (if self--gravitating) $M_{\rm
dyn} \sim 3 \times 10^9\,M_\odot$ at least, or higher according to its inclination,
 roughly coincident with the location of a peak in  8.4\,GHz continuum.

\item The morphologies of the molecular gas structures are quite asymmetric, 
with a one--armed spiral winding through roughly $270^\circ$ in 
azimuth, or more likely (since the pitch angle is close to zero) 
a molecular ring.
The CO(1--0) reveals a ring of diameter 35\arcsec = 8.9\,kpc, and the
CO(2--1) reveals an asymmetrically embedded  ring of 
diameter 9\arcsec = 2.3\,kpc. A hint of a nuclear gas
bar is visible in the high-resolution CO(2--1) map, with a diameter of 
0.7\,kpc,  and a PA of 15$^\circ$.
There is a nuclear stellar bar of roughly 1\,kpc in diameter,
revealed by {\it HST} in the near-infrared, with a
PA of -50$^\circ$.
  We note that the two CO rings at 
radii of roughly 1 kpc and 4 kpc could correspond
to the ILR and OLR of this bar, if it had a
pattern speed of $\sim 120\,{\rm km\,s^{-1}\,kpc^{-1}}$ 
  (from Fig. \ref{vcur}) .

\item The southwestern CO(1--0) emission coincides with a peak in
the near infrared map, at about 5\,kpc from the center of NGC\,1961.
 This NIR secondary peak is split in three smaller  components, 
suggesting the disruption of a stellar system. This splitting
could also be partially due to dust, although not likeky, given
the moderate average gas column density in the region. 

\end{enumerate}
We can use these facts to help evaluate the merits of the minor merger and 
ram pressure stripping scenarios which have been proposed to account for the 
large--scale asymmetries seen in NGC\,1961.  An initial, obvious point is that 
because these asymmetries are seen in stars as well as in gas (e.g., Figure 
\ref{f-opt}), the classical version of ram pressure stripping conceived as 
acting only upon gas (Gunn \& Gott \cite{gunn72}) will not be sufficient.  A 
more nuanced stripping scenario would resemble the recent simulations of 
Vollmer et al.  (\cite{voll01})  and Vollmer (\cite{voll03}), who have 
shown that it is possible for a galaxy's 
stellar potential to {\it respond} to a strong perturbation induced by the gas 
ram pressure.  However, in the case of NGC\,1961, there appears to be too 
little intragroup medium to allow for this.  
If there {\it were} sufficient hot IGM, the geometry of the H\,I tail would
correspond nicely to what is expected from ram pressure.
The buildup of HI and enhancement of radio continuum emission in
NGC\,1961's southeastern ridge could then be accounted for by an edge-on
stripping geometry. For such a geometry, simulations predict 
 that a galaxy encountering an 
intragroup medium for the first time should have its H\,I extended on the side 
where the galactic rotation adds to the velocity relative to the IGM 
and the gas is more stripped on the other side
(see the explanation of this paradox in  Roediger \& Br\"uggen  \cite{roed06}; 
Jachym et al.  \cite{jach07}, \cite{jach09}).  For NGC\,1961, 
which is falling into its group from the northwest, we see an extended H\,I 
structure on the northern side of the galaxy, coming from the west (Haan et al. 
\cite{haan08}). For ram pressure to produce this configuration,
the velocity of NGC\,1961 relative to its group should be blueshifted
towards us (i.e. the galaxy should be located behind the main group). 
Thus the ram-pressure scenario is plausible from a geometrical point of view,
the only problem is the insufficient IGM.

In the absence of IGM, we arrive therefore at the conclusion that a tidal 
encounter or minor merger is the most likely explanation for NGC\,1961's 
current state of disruption.  The galaxy interaction cannot
be a prograde grazing passage,  as in this case extended tidal H\,I tails will form,
which are not observed. The interaction must be either retrograde,
or nearly head-on. The rings we have detected in the CO maps
favor a  head-on encounter, with a small impact parameter.

Since there is no obvious interaction partner remaining on the scene, it is also reasonable 
to conclude that the smaller galaxy has already merged with NGC\,1961; this hypothesis has 
the advantage of naturally accounting for the distinct kinematic axis of
the southwestern peak in CO(1--0).  We can 
expect {\it a priori} that the merger mass ratio should be of order 1:5, to 
account for the substantial (but not total) disruption of the original disk
of the smaller galaxy.
These considerations together provide us with starting conditions  to develop a 
detailed minor merger model.

The {\it Spitzer} IRAC images bring  new insight on the true morphology of NGC\,1961.
As can be seen in Figure \ref{nonstar}, the dust emission clearly reveals 
two off-centered rings, in a morphology similar to that of the Cartwheel galaxy
(e.g. Appleton \& Struck-Marcell \cite{appl96}).
In particular, the ring morphology is closely related to spokes in the disk:
multiple spokes link the inner ring to the outer off-centered ring,
starting nearly tangential to the inner ring and ending perpendicular 
to the outer ring.  These specific details are not so apparent
in the optical image, which is confused by star-formation spots and extinction,
although the spokes still can be seen.
 The spokes bridging the gap between off-centered rings are typical
of a head-on collision in a spiral galaxy: either the latter possesses a spiral structure 
with a large pitch angle before the collision, and this structure is superposed
on rings that propagate outwards from the center during the collision
(e.g., Block et al. \cite{bloc06}), or the spiral structure forms through
gas instabilities during the collision (e.g., Hernquist \& Weil \cite{hern93}). 

\begin{figure}
\centering
\includegraphics[width=8.5cm]{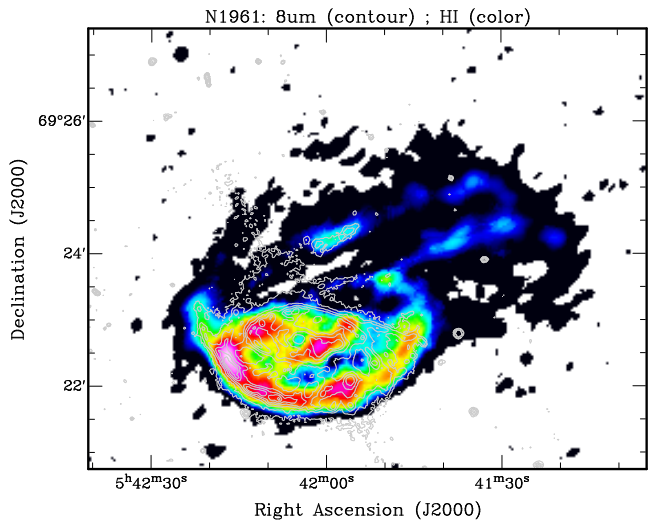}
\caption{The {\it Spitzer} 8\,$\mu$m contours are superposed on the 
color map of the H\,I surface density from Haan et al  (\cite{haan08}). }
\label{HI-8mu}
\end{figure}

 The  ring and spokes morphology  of even the central stellar and ISM components
in NGC\,1961 suggests a nearly head-on encounter.  Of course,  this could also be superposed
on some ram-pressure phenomenon. However, the superposition of the H\,I
and IRAC PAH-dust images in Figure \ref{HI-8mu} shows a rather good correspondence,
suggesting a dominant gravitational phenomenon, which will similarly drag the different
ISM components, diffuse or clumpy. The stellar component, traced by the IRAC 3.6$\mu$m
image, also follows the sharp southern arc (Pahre et al. \cite{pahr04}).

\begin{figure}
\centering
\includegraphics[width=8.5cm]{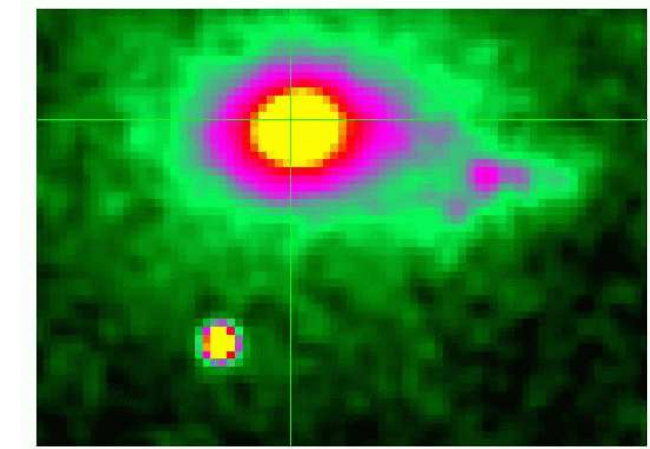}
\caption{2MASS $K$-band image of the central 76.6\arcsec $\times$ 55\arcsec
of NGC\,1961. The cross-hair indicates our adopted phase
center for the PdBI observations (the SE star is at 29.6\arcsec from
this center, and can serve as a reference for the resolution of the image. 
North is up, and east is to the left).
The NIR images all reveal a high stellar density at about 24\arcsec
to the southwest, which is split in three pieces. This location corresponds
to the high velocity gradient detected in the CO(1-0) map, and suggests the presence 
of a small disrupting satellite.}
\label{2mass-k}
\end{figure}

The possible existence of the satellite remnant, suggested already by
Lisenfeld et al. (\cite{lise98}) from the radio map, is also supported
by the near-infrared images that trace the stellar density. In the 2MASS
images (see Figure \ref{2mass-k}), there is clearly a second nucleus,
which appears split into 3 clumps. Its average distance from the 
center of NGC\,1961 is 24 \arcsec or 6.1 kpc.
This feature cannot be seen in the {\it HST} NICMOS images, since
it falls outside of the field of view. 

The NOT image of NGC\,1961\footnote{see the web site
http://www.not.iac.es/general/photos/ astronomical/extragalactic/}
constructed by J. N\"ar\"anen \& K. Torstensson in 2004 by combining $B$, $V$,
and $R$ exposures shows that the satellite nucleus is partially hidden
by a dust lane, and is split into 2--3 clumps.  It is clear in this image 
that star formation is strongly enhanced in the rings,
which are delineated by multiple blue hot spots.

Summarizing, the evidence that NGC\,1961 has experienced a head-on collision 
followed by a minor merger includes:
\begin{enumerate}

\item  the off-centered ring-like morphology (recalling that the main arcs/arms
have no obvious sense of winding), observed in many components and 
wavelengths, including in CO and in dust emission (e.g., Fig. \ref{nonstar});

\item  the very sharp southern edge of the disk observed in the gas and dust component,
in particular in the H\,I 21\,cm map (see Figure \ref{HI-8mu}), which could 
be the tracer of an expanding ring wave;

\item  the existence of ``spokes'' linking the ring features, with a large
pitch angle, particularly visible
in the Spitzer dust image (Figure \ref{nonstar}); and 

\item  the presence of a perturbation in the CO
velocity field 5\,kpc from the center, coinciding with a secondary stellar 
peak and a radio continuum peak.

\end{enumerate}

\section{Simulation of the head-on collision}
\label{simu}

 To test the plausibility of a head-on encounter leading to a minor merger
in NGC\,1961, we performed a self-consistent simulation of the collision.

\subsection{Galaxy model and numerical techniques 
\label{code}}

The simulations are 3D $N$-body with stars and gas but 
not including star formation.
They are fully self-consistent, with a live dark halo, and 
use a Particle-Mesh code based on FFT with a useful grid of
128$^3$ (the algorithm of James \cite{jame77} allows us to suppress
the Fourier images).
The grid size is (72\,kpc)$^3$, and the model galaxy is truncated
at an initial radius of 22\,kpc.
The softening is equal to the size of a cell, i.e., 560\,pc.
The time step is 1\,Myr.
 The gas is represented by sticky particles,
and a total of 2.4 10$^6$ particles are used.
The non-dissipative components, stars and dark matter,
are represented by 2 million particles, and the gas component by 4 10$^5$.
  First tests for the simulations were done with 10 times less particles,
and the total number were used only when a reasonable fit was found.
The comparison between simulations  with different particle numbers
reveal that the main features (propagating ring waves, off-centering) are a fundamental
and persisting result, while secondary ones, like the exact positions of spokes, 
or the contrast of the arms/rings, are changing. 
 
The stellar component is composed of a bulge and a disk.
The bulge is initially distributed as a Plummer sphere, with a potential:
\begin{equation}
\Phi_{b}(r) = - { {G M_{b}}\over {\sqrt{r^2 +r_{b}^2}} }
\end{equation}
where  $M_{b}$ and  $r_{b}$ are the mass and characteristic radius of the
bulge, respectively (see Table \ref{condini}). 

The stellar disk is initially a Kuzmin-Toomre disk of surface density
\begin{equation}
\Sigma(r) = \Sigma_0 ( 1 +r^2/r_d^2 )^{-3/2}
\end{equation}
with a mass  $M_d$,
and characteristic radius $r_d$.
It is initially quite cold, with a Toomre $Q$ parameter of 1.
The dark matter halo is also a Plummer sphere, with mass $M_{\rm DM}$ and characteristic radius
$r_{\rm DM}$.  The initial conditions  of the run
described here are given in Table \ref{condini}.

\begin{table}[ht]
\caption[ ]{Initial conditions}
\begin{flushleft}
\begin{tabular}{ccccccc}  \hline
Component       & radius & Mass & Mass fraction   \\ 
                          & [kpc]  & [$M_\odot$]&    [ \%]                \\
\hline
Bulge          &  0.5    &  4.e10      &  13.3       \\
Disk             &  6.5   &  9.e10      &   30.       \\
Halo$^*$    &  15.   & 15.e10    &   50.     \\
Gas              &  8.      &  2.e10     &  6.6         \\
Companion & 2.5    &   7.e10  &  23.3     \\
\hline
\end{tabular}
\end{flushleft}
$^*$ Halo mass inside 22\,kpc radius\\
The mass fractions are normalized to the total mass of NGC\,1961,  $3\times 
10^{11}\,M_\odot$ inside 22\,kpc\\
\label{condini}
\end{table}

The gas is treated as a self-gravitating component in the $N$-body
simulation, and its dissipation is represented by a sticky particle code,
following Combes \& Gerin (1985). We varied the  initial gas-to-total mass ratio 
between 4 and 8\% to control the stability of the disk prior
to the collision.
 The mass of one gas particle therefore varied between $3 \times 10^4$ and 
$6 \times 10^4\,M_\odot$.
  This is meant to represent essentially the molecular gas.
 The initial gas distribution  is also  a Kuzmin-Toomre disk,
but truncated at 34\,kpc, and with a characteristic radial scale of 8\,kpc.
Initially, its velocity dispersion corresponds to a Toomre $Q$ parameter
of 1. The gas clouds are subject to inelastic collisions, with a collision
cell size of 560\,pc (the volume within which particles are selected
to possibly collide). This corresponds to a lower limit for the average mean
 free path of clouds between two collisions. The collisions are considered
every  10 Myr. In a collision, the sign of the relative
cloud velocities is reversed and the absolute values are reduced: 
relative velocities after the collision are only $f_{el}$ times their 
original value, with the elasticity factor $f_{el}$ in the radial direction 
controlling the dissipation rate. The $f_{el}$ 
parameter must be adapted to the particle number, so that a given
gas particle experiences a reasonable loss of energy in a rotation. 
This is calibrated in simulations of the isolated galaxy, where there
is an equilibrium between the dissipative cooling,
and the dynamical heating due to spiral arm formation.
The $f_{el}$  parameter 
has been fixed to 0.65 for low particle numbers, and 0.85 for high particle
numbers (10 times more particles).
The analogous coefficient in the tangential
direction is fixed to 1, to conserve angular momentum in each
individual cloud collision.
All gas particles have the same mass ($5 \times 10^4\,M_\odot$ in the
standard model). The particle mass for the dissipationless 
components, stars and dark matter, is identically 
$1.5 \times 10^5\,M_\odot$.

\begin{figure}[ht]
\rotatebox{-90}{\includegraphics[width=6cm]{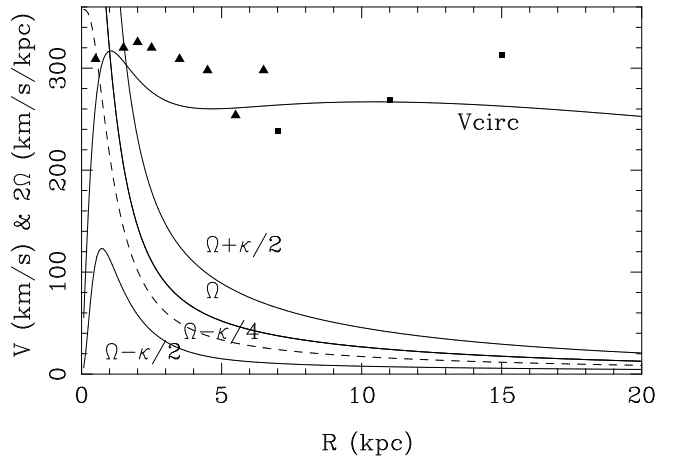}}
\caption{ Rotation curve and derived frequencies $\Omega$,
$\Omega-\kappa/2$ and $\Omega+\kappa/2$ adopted 
for the simulations presented here. The triangles are the CO
rotational velocities; the squares are from the H\,I data (Haan et al 2008),
with an adopted inclination of the galaxy main plane of 65$^\circ$.}
\label{vcur}
\end{figure}

The rotation curve corresponding to the standard run
is plotted in comparison to the data points in Fig.~\ref{vcur}.
All other runs have similar rotation curves.
 The parameters varied were essentially the geometry
of the interaction, the mass of the companion, and the stability of the disk before
the interaction.

 One of the most surprising characteristics of NGC\,1961 is its high dynamical
mass relative to its late type. This problem is alleviated by the choice
of a higher inclination of the main disk with respect to the plane of the sky.
It will be shown in the next section that the outer disk is warped, making
the galaxy artificially appear more face-on. Here we adopt a high inclination
of 65-70$^\circ$ for the initial disk orientation,  and hence for the 
calculated rotation curve.

The companion is represented by a Plummer potential,
where radius and mass are 
given in Table \ref{condini}. The collision corresponds 
to a minor merger with a mass ratio around 1:4.
Since we want to explain the morphology of NGC\,1961, and
have no constraint on its satellite,  the companion is simulated
as a rigid body. Its orbit is computed from the forces 
exerted on the satellite by all the particles in the main galaxy, so that most
of the dynamical friction is taken into account.  The contribution
to dynamical friction from the companion disruption is not considered,
and could decrease the merging time  (see Prugniel \& Combes \cite{prug92}).
 Given the many free parameters, we do not aim to reproduce exactly 
the whole collision or the exact position of the companion now,
but just to demonstrate that this type of collision is able to provide the 
right order of magnitude for the perturbation, and morphologies similar 
to those observed.

\subsection{Simulation results}
\label{sim-res}

 About a dozen runs have been carried out, to test the various geometrical parameters
of the encounter and also to select the initial morphology of the galaxy.
Indeed, a bar and a spiral wave develop  in the disk during the first
period when the galaxy simulation is run in isolation, and the 
starting time of the collision plays an
important role in determining the final morphology. 
Similar experiments were carried out to simulate the head-on
collision of the Andromeda galaxy: a small satellite (possibly M32)
could have triggered the propagation of two ring waves in the disk, which 
superposed on the pre-existing spiral structure give the complex
and perturbed morphology of M31 (cf Block et al \cite{bloc06}).

The orbit of the companion is defined by its energy and its impact
parameter. 
 The energy is computed from the equivalent two-body problem,
assuming that all the mass of NGC\,1961 is concentrated in a point
(of mass $M$).
The energy per unit mass of the reduced particle is then:
\begin{equation}
E = \frac{1}{2} V^2 - G M (1 + \mu)/R
\end{equation}
where $\mu$ is the mass ratio of the satellite to the main galaxy, 
and $R$ and $V$ are the relative distance and velocity between
the two.
 
The initial distance of the companion is 70\,kpc, with a dominant
component  perpendicular to the plane of the main galaxy.
 The kinetic energy and relative velocity are then obtained, assuming 
either a ``hyperbolic'' type of orbit with positive energy,
or an ``elliptic''  type with negative energy.  The dynamical friction
then reduces the relative energy from its initial value, and all orbits
end up as bound. The best run (presented
here) has initially a negative energy, of $-0.87 \times 10^4\,{\rm 
km^2\,s^{-2}}$.
 
The various experiments revealed that the 
impact parameter must lie between 4 and 10\,kpc; it can be
no larger than 10\,kpc if the ring waves characteristic
of a head-on collision are to appear, but no smaller than 4\,kpc if the 
collision is to generate sufficient asymmetry and off-centering of the rings.

\bigskip

During the head-on collision, part of the outer disk is raised outside
the plane in a huge warp. This global motion should be even
more visible in the outer gas disk, and we suggest that the conspicuous
northwestern features seen in the H\,I  map, forming
 a fan with low-projected radial velocity along the line of sight,
is a consequence of this warp. The deformation of the plane
to a more face-on orientation has thus deceived us as to 
the true inclination of the main plane and the inferred
dynamical mass.

\begin{figure}
\includegraphics[angle=-90,width=8.5cm]{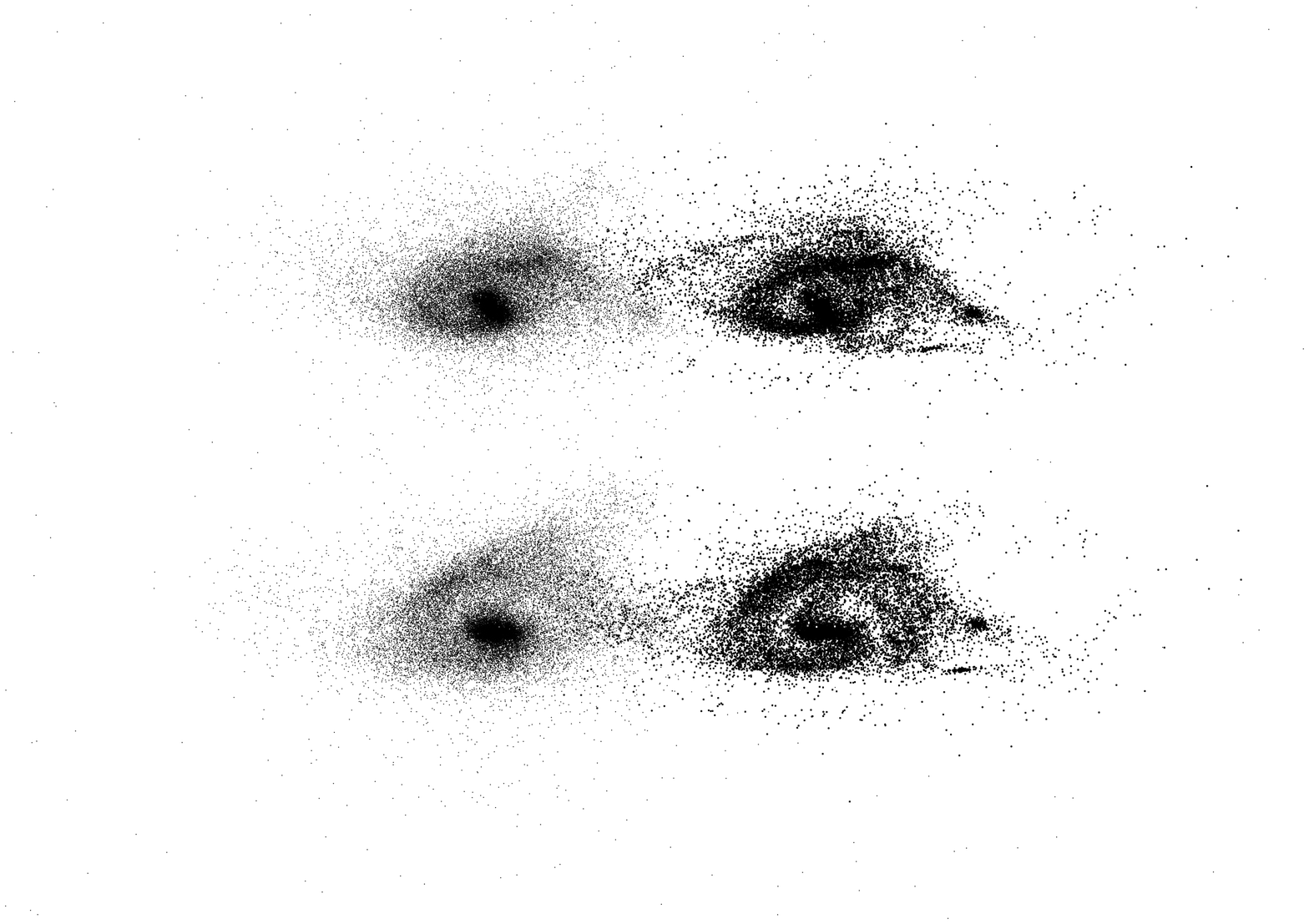}
\caption{ Particle plots from the simulation of the collision,
projected to be similar to the observations (inclination 70$^\circ$):
{\bf Left:} the stellar component, {\bf Right:} the gas component.
 The two snapshots correspond to 320 and 360 Myr (top and bottom 
respectively). The size of each projected field is 63 kpc.}
\label{simplot}
\end{figure}

\begin{figure}
\includegraphics[angle=-90,width=8.5cm]{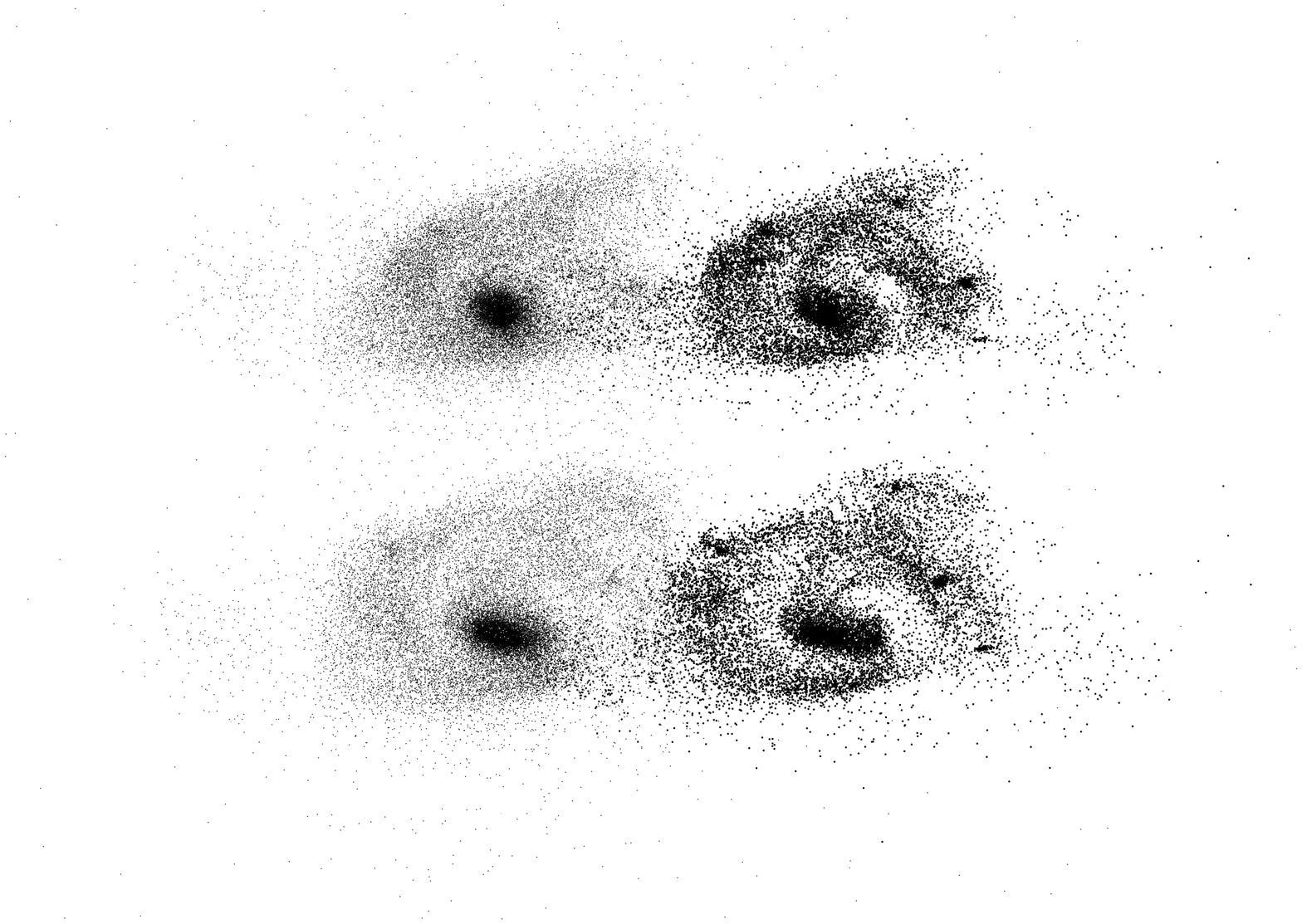}
\caption{ Same as Fig.~\ref{simplot}, but now the 
snapshots correspond to 400 and 440 Myr (top and bottom 
respectively). Note how the radius of the galaxy inflates while the 
ring wave propagates outward.}
\label{simplot2}
\end{figure}

In the run presented in a few snapshots in
Fig. \ref{simplot} and \ref{simplot2}, the initial galaxy model
was run in isolation for 280 Myr, and formed a barred spiral.
 Then the companion was introduced at 70 kpc, and the collision
was followed during the subsequent 440 Myr. 
 At this epoch, estimated to give the best fit to the
observations, the ring waves have already begun to propagate outwards,
and it is easily seen how the apparent radius of the galaxy inflates
in the figures.
After 1.5 Gyr of evolution, the galaxy could then become a giant LSB galaxy,
as proposed by Mapelli et al. (\cite{mape08}).

\begin{figure}[ht]
\rotatebox{0}{\includegraphics[width=8.5cm]{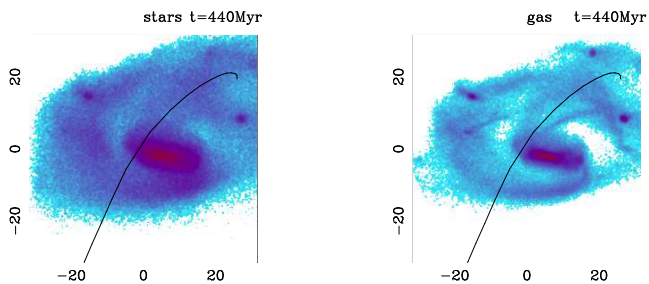}}
\caption{Projected density of the model, at a time of 440 Myr after
the beginning of the collision: the stars {\it Left} , and the gas {\it Right}.
The scale is in kpc. The black line is the trajectory of the companion
starting from the southeast.} 
\label{run4-440}
\end{figure}

\begin{figure}[ht]
\rotatebox{0}{\includegraphics[width=8.5cm]{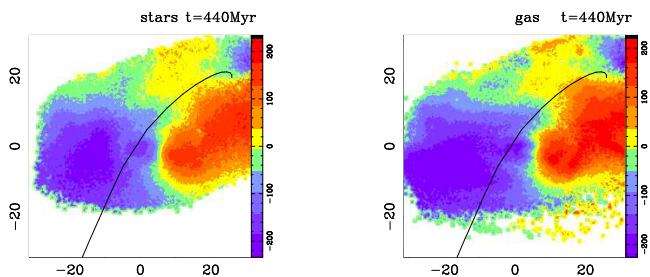}}
\caption{ Same as Figure \ref{run4-440}, for  radial velocities.
 The color bar indicated at the right side goes from $-210$ to $210\,{\rm 
km\,s^{-1}}$.} 
\label{run4-vel}
\end{figure}

A density map of the stellar and gas component 
is plotted in Figure \ref{run4-440} and the corresponding 
velocity fields in Figure \ref{run4-vel}.
It is possible to see a small accumulation of particles accreted
by the companion on its way back to merge with the main galaxy.
  Since the companion's position is not fit to what is observed, the 
velocity gradient observed around it is located too far north; however,
the physics of the satellite were not actually addressed in these
simulations. It is not even known whether the molecular gas
observed in this region is coming from the main galaxy or from
the merging satellite.

The velocity fields do show the  gross features of what is observed,
to order of magnitude. In particular, they show reasonably regular rotation
in spite of a quite perturbed galaxy ``plane''.

A feature that is well reproduced by the collisional model is
the sharp boundary in the southern edge of the gas disk
(Figure \ref{run4-440}). This sharp edge was one piece of evidence that 
had seemed to favor the ram pressure shock scenario. An additional point of 
agreement 
is that the gas boundary appears in projection to be shifted to the north
with respect to the stellar component boundary. This is because 
 the gas disk is much thinner than the stellar disk, which is 
heated in the collision. The fact that the gas front was pushed ``inside''
the stellar disk was advanced as one of the main arguments to suggest
a ram-pressure scenario by Shostak et al. (\cite{shos82}).
In the present collisional model, it is only a projection effect, arising from
two components of different geometrical thickness.

Another feature reproduced by the simulations is the "spoke" morphology 
connecting the two off-centered rings (see Sect. 4).  This kind of feature is
typical for head-on collisions, as is shown in the Cartwheel simulations
(e.g. Hernquist \& Weil, 1993; Horellou \& Combes, 2001). These are visible in 
the models in the gas component only, similar to the 8-micron dust-only 
image where they are clearly visible. Although the spokes in NGC 1961 are 
not at the same locations as those in our simulations, the similarity suggests that a 
head-on collision is capable of reproducing, at least qualitatively, the 
main morphological features observed.

\section{Discussion}
\label{discuss}

The head-on collision model reproduces quite well the ring
morphology and the sharp gas edge, which were difficult to 
account for with ram pressure, given that hot IGM is almost absent
in this small group of spiral galaxies. The 
minor merger scenario can also hold 
in the absence of a nearby interacting galaxy. 

One interesting outcome of the simulation is that it also 
gives a possible solution to the surprisingly large radius of the galaxy
with respect to its late (Sc) Hubble type.  The outward propagating
ring waves of the collision can increase the disk size to what appears
to be abnormal proportions.  

One question that can still be asked is how such a galaxy in a relatively
rich environment can retain such a late type, without a significant
bulge.   It may be that the galaxy is entering the group for the first time.
Alternatively,  even if it has already interacted and merged with some members
of the group, they are all gas rich (Shostak et al. \cite{shos82}) and  
could have replenished the disk. The in--plane accretion of a 
large number of satellite galaxies and intergalactic H\,I clouds with much 
lower masses can quietly build up the disk mass without dynamically heating a 
spheroid (see Bournaud \& Combes \cite{bour02}), thereby depressing NGC\,1961's
bulge--to--disk ratio and driving its evolution towards later Hubble type.  
The most recent encounter has been sufficiently disruptive, however, that we 
would predict a reversion to an earlier type as the result of dynamical heating.
The fact that we do not see {\it more} galaxies which are too massive for 
their Hubble types, suggests that this merger of moderate mass ratios 
must correspond to a quick transient phase in poor groups.

NGC\,1961 is a low-luminosity AGN. From the central velocity dispersion,
the mass of the central black hole can be estimated to be
$3 \times 10^8\,M_\odot$ (Krips et al. \cite{krip07}). From the X-ray 
luminosity of the nucleus and its deduced bolometric luminosity, its 
Eddington ratio is estimated to be $L_{\rm bol}/L_{\rm Edd} 10^{-6} - 10^{-5}$.
 This means that the fuelling is not intense at this moment, and  must be
intermittent. The very different time-scales  between dynamical phenomena
at the extreme large and small scales is indeed an issue.

NGC\,1961 is however a good candidate for minor-merger induced
fuelling of a low luminosity AGN; at least the minor merger made more
potential fuel available in the inner kpc.
This was proposed by several authors (see e.g., Taniguchi \cite{tani99}),
and debated by others (e.g., Corbin \cite{corb00}).
Hopkins \& Hernquist (\cite{hopk09}) recently proposed to separate the
fuelling of high-luminosity AGN (quasars), which would be fuelled by major mergers,
from the  low-luminosity ones, where secular evolution or minor merging
is sufficient. NGC\,1961 could be just an example of the transition between
the two regimes.

\section{Conclusions} \label{s-conc}

Based on our observational and theoretical investigation of the asymmetries 
evident in NGC\,1961 on both large and small scales, we have reached the 
following conclusions:
\begin{enumerate}

\item The galaxy's large-scale optical asymmetries can be explained as the 
result of a head-on collision followed by a minor merger with mass ratio 
of $\sim$ 1:4.   Models that rely 
exclusively on ram pressure stripping cannot account for the 
level of disturbance observed in the galaxy's stellar light.

\item Our preferred minor merger model can naturally account for the 
excitation of $m = 1$ asymmetries in the galaxy's inner disk, and also 
for embedded off-centered rings  in the CO maps. It can also explain the 
very sharp gas boundary in the southern part.  Due to the geometrically thin gas
plane, this sharp edge appears in projection shifted inside the stellar disk, a
feature that had previously been thought to favor only a ram-pressure scenario.

\item The head-on collision explains the ring wave propagating outwards,
but also the spokes with large pitch angle that connect the two rings.
On the smallest scales, the relaxation time is sufficiently small compared to
the merging timescale that the inner molecular disk has been able to become
bar--unstable, as reflected in the $m = 2$ or nuclear bar morphology seen in 
the CO(2--1) map.
The galaxy could have been barred before the merger, but the latter could
have enhanced the bar strength, or create a new one, 
helping the gas inflow to fuel the AGN. 

\item Our collision scenario explains the exceptionally large
size of NGC\,1961 for its late type. The large disk warp accounts
for an under-estimation of its true inclination in the plane of the sky,
and thus an over-estimation of its dynamical mass.  NGC\,1961 
may have built up a significant fraction of its current 
disk mass by the relatively quiescent accretion of intracluster gas.  This 
scenario can account for the low bulge-to-disk ratio of this spiral galaxy in a 
relatively rich environment.  The most recent and traumatic 
merger event, however, is likely to force NGC\,1961 to revert to an earlier 
Hubble type.

\end{enumerate}

The simulations presented here only demonstrate the plausibility of the head-on 
collision, and are not intended to provide an exact fit to the observations. In particular,
the companion is only schematized as a rigid body,  since there are insufficient 
observational constraints to define its physical parameters.
 We conclude however that  NGC\,1961  could be a representative example of a minor-merger
induced low-luminosity AGN.

\begin{acknowledgements}
We are grateful to IRAM staff members at the PdBI and in Grenoble for making 
these observations possible.  We thank Ute Lisenfeld for providing electronic 
versions of her radio continuum maps. This
work made use of the HyperLEDA and NED databases.

\end{acknowledgements}

\end{document}